\documentclass[]{aa}
\usepackage{graphicx}
\usepackage{txfonts}
\begin{document}
       \title{Search for radiative pumping lines of OH masers:}

       \subtitle{I. The 34.6\,$\mu$m absorption line towards
1612\,MHz OH maser sources.\thanks{Based on observations with ISO, an ESA project with
instruments funded by ESA Member States (especially the PI countries: France,
Germany, the Netherlands and the United Kingdom) and with the participation of
ISAS and NASA.}
	           }

       \author{J.H. He\inst{1}
              \and R. Szczerba\inst{2}
              \and P.S. Chen\inst{1}
              \and A.M. Sobolev\inst{3}
              }

       \offprints{J.H. He, \\ \email{mailhejh@yahoo.com.cn}}

       \institute{National Astronomical Observatories/Yunnan Observatory,
	            CAS, P.O. Box 110, Kunming, 650011, China
             \and
                 N. Copernicus Astronomical Center, Rabia\'{n}ska 8,
                 87-100 Toru\'{n}, Poland
             \and
                 Astronomical Observatory, Ural State University, Lenin Street 51,
                 Ekaterinburg 620083, Russia
                 }

          \date{Received / Accepted }

       \abstract{
       The 1612 MHz hydroxyl maser in circumstellar envelopes has long
been thought to be pumped by 34.6\,$\mu$m photons. Only recently, the Infrared 
Space Observatory has made possible spectroscopic observations which enable
the direct confirmation of this pumping mechanism in a few cases. To look 
for the presence of this pumping line, we have searched the Infrared Space 
Observatory Data Archive and found 178 spectra with data around 34.6\,$\mu$m for 87 
galactic 1612\,MHz masers. The analysis performed showed that the noise level and the
spectral resolution of the spectra are the most important factors affecting the 
detection of the 34.6\,$\mu$m absorption line. Only 5 objects from the sample 
(3 red supergiants and 2 galactic center sources) are found to show 
clear 34.6\,$\mu$m absorption (all of them already known) while two additional
objects only tentatively show this line. The 3 supergiants show similar pump 
rates and their masers might be purely radiatively pumped.
The pump rates of OH masers in late type stars are found to be about 0.05,
only 1/5 of the theoretical value of 0.25 derived by Elitzur\,(\cite{eli92}).
We have also found 16 maser sources which, according to the analysis assuming
Elitzur's pump rate,  
should show the 34.6\,$\mu$m absorption line but do not. These non-detections 
can be tentatively explained by far-infrared photon pumping, clumpy nature of the 
OH masing region or a limb-filling emission effect in the OH shell. 

	     }

	\titlerunning{$34.6\mu$m radiative pumping line of 1612\,MHz OH maser}
	\authorrunning{He et al.}

       \maketitle

%
%________________________________________________________________

\section{Introduction}

The first detection of intense radio emission from OH molecules was reported by
Weaver et al.\,(\cite{wea65}) and soon an explanation based on maser
amplification through induced processes was invoked (Litvak et
al.\,\cite{lit66}, Perkins et al.\,\cite{per66}). Shklovsky\,(\cite{shk66})
was the first to propose a radiative pumping mechanism for OH masers.
Harvey et al.\,(\cite{har74}) suggested that the correlation between the infrared (IR)
and 1612\,MHz variabilities observed in IR stars that exhibit this satellite
line maser (hereafter OH/IR stars) is probably due to a radiative coupling
mechanism between the stars and the OH clouds (radiative pumping of the maser,
possibly at 2.8\,$\mu$m or 34.6\,$\mu$m). The idea of radiative pumping for
OH/IR stars was then elaborated in detail by Elitzur et
al.\,(\cite{eli76}), who concluded that a pumping mechanism by far-infrared
photons is more compatible with the observations. In this model, the required
inversion of $F$=1 and $F$=2 sub-levels (even and odd parity, respectively) in
the lowest rotational level of the OH molecule is achieved by absorption of
infrared photons at 34.6\,$\mu$m from the ground state ($^2{\Pi}_{3/2}\,J$=3/2)
and consequent radiative decays to lower levels via other far-infrared
transitions. The final and the most crucial step in the pump cycle is the
radiative decay from $^2{\Pi}_{1/2}\,J$=1/2 to  $^2{\Pi}_{3/2}\,J$=3/2.
If the transitions that link the two involved rotational level ladders are
optically thick, a strong inversion of the level population that gives rise to the
1612\,MHz maser can be produced. However, the fact that the 34.6\,$\mu$m photons 
alone are enough to pump the 1612\,MHz masers in OH/IR stars does not exclude that
some other factors, such as collisional effects, line overlap, near infrared (NIR) 
pumping, might affect the maser pumping process. Elitzur et al.\,(\cite{eli76}) 
discussed these factors 
but did not include them in their model. Under the assumptions of optically thick 
far-infrared (FIR) 
pumping lines and saturated OH masers, their model showed that four FIR 
photons at 34.6\,$\mu$m are needed to produce one maser photon at 1612\,MHz.
Bujarrabal et al.\,(\cite{buj80}) improved the OH 18\,cm maser pumping model for 
circumstellar envelopes by adding line overlap effects into the model. Their modeling work 
showed that FIR radiation together with the FIR line overlaps 
dominate the pumping of both OH main line and  satellite line masers and the FIR line overlaps 
can strengthen the 1612\,MHz maser. In order to explain the intensities of the main line OH masers 
observed in OH/IR stars and type II Miras, Collison and Nedoluha\,(\cite{col94}) included 
additional processes in their OH maser model: non-local FIR line overlaps, collisional 
excitations, dust FIR pumping. They also investigated the importance of NIR 
line overlap and pointed out that the NIR line  overlap between OH and H$_{2}$O can be 
important for the inversion of main line OH masers in thinner envelopes. 
Thai-Q-Tang et al.\,(\cite{tha98}) also developed an OH maser model to reproduce 
the observed FIR OH absorption lines in a red supergiant (RSG) OH maser: \object{IRC+10420}. 
Their model included collisional and FIR pumping but the line overlaps were only considered 
among the microwave transitions in the ground rotational level. They found that the observed 
maser properties are better reproduced if the Doppler shift is confined to a small range 
($\sim$2 km\,s$^{-1}$), and argued that other processes such as a clumpy nature 
of the OH shell, NIR line 
overlaps or local thermal line overlaps may also play important roles in the pumping of the OH 
masers. Conclusively, we can say that problems related to the OH maser pumping are still poorly 
understood and therefore are worthy of further efforts. One of the main purposes of this 
paper is to search for more evidence of the FIR radiative pump mechanism in OH
1612\,MHz maser sources.

Although the radiative pumping mechanism for the stellar 1612\,MHz OH masers by the 34.6\,$\mu$m
line has been proposed long ago, its
direct confirmation by observations became possible only after the
launch of the Infrared Space Observatory (ISO, see Kessler
et al.\,\cite{kes96}). The 34.6\,$\mu$m line was found in the red supergiant
\object{NML\,Cyg} by Justtanont et al.\,(\cite{jus96}), and in the 
ISO spectrum obtained with the aperture centered on the galactic center 
object \object{Sgr\,A*} by Lutz et al.\,(\cite{lut96}). Then,  Sylvester et
al.\,(\cite{syl97})  found it together with the other OH rotational cascade
emission lines at 79, 98.7 and 163\,$\mu$m towards \object{IRC$+$10420},
another well known supergiant. The 1612\,MHz 
OH maser in the envelope of this star was modeled by Thai-Q-Tang et
al.\,(\cite{tha98}) who showed that these observations confirmed the radiative 
pumping of the maser by 34.6\,$\mu$m line radiation. The 34.6\,$\mu$m
absorption feature was also
reported by Neufeld et al.\,(\cite{neu99}) for another supergiant,
\object{VY\,CMa}, and by  Goicoechea \& Cernicharo\,(\cite{goi02})
{in the ISO spectra taken towards another galactic center object}
\object{Sgr\,B2}. Note that the 34.6\,$\mu$m absorption feature has also
been found in ISO spectra associated with four megamasers: the famous
interacting galaxy \object{Arp\,220} (Skinner et al.\,\cite{ski97}), the
starburst galaxy \object{NGC\,253} (Bradford et al.\,\cite{bra99}),
\object{IRAS\,20100$-$4156} and probably \object{III\,Zw\,35} (Kegel et
al.\cite{keg99}). All these galaxies are predominantly 1667 MHz main line OH
masers and in such cases the radiative pumping mechanism requires additional
conditions to be fulfilled (see e.g. Elitzur\,\cite{eli78}, Bujarrabal et
al.\,\cite{buj80} or Elitzur\,\cite{eli92} for more detailed discussion). In
this paper we concentrate on the galactic OH masers in the satellite line at
1612\,MHz.

The discoveries mentioned above call for a more systematic check of the
expected general presence of the 34.6\,$\mu$m OH absorption line in other ISO
spectra that are associated with the galactic OH-IR objects (i.e. OH/IR stars and non-stellar
IR sources with OH masers). We have searched for the 34.6\,$\mu$m 
feature in the ISO database for all 1612 MHz OH-IR objects irrespective 
of their evolutionary status (i.e. AGB stars, young stellar objects, 
molecular clouds, etc.) to check the frequency of the detection of this line. 
A preliminary analysis of
available ISO data around 34.6\,$\mu$m for OH-IR objects from the Chen et
al.\,(\cite{che01}) catalog has been presented by Szczerba et
al.\,(\cite{szc03}). Here we present an improved analysis of 178 ISO Short Wavelength
Spectrometer (SWS - de Graauw et al.\,\cite{deg96}) spectra obtained towards 87
galactic OH-IR objects with the aim to search for the 34.6\,$\mu$m OH
absorption line. In Section\,\ref{obsprocess} we describe our sample of
galactic OH-IR objects and the data processing. The analysis of the
detection rate of the 34.6\,$\mu$m line is given in
Section\,\ref{detectionrate} and is followed by the discussion of pump rate for
sources with detection and possible explanations for the absence of the pumping
line in sources with non-detection of the 34.6\,$\mu$m absorption line in
Section\,\ref{OHpump}. A summary is presented in Section~\ref{summary}. 
In future papers we will analyze the 
available ISO data to search for other OH rotational absorption 
and cascade emission lines.

\section{Observations and data processing}
\label{obsprocess}

The present version of Chen's catalog (unpublished) of 1612\,MHz OH maser
sources contains 1940 galactic objects. These OH sources were cross-correlated
with the IRAS Point Source Catalog (PSC). Out of the 1940 entries, 1876
objects have their maser positions not farther than 1\arcmin\ from
the positions of the nearest IRAS PSC object and they constitute our working
sample of galactic OH-IR objects in this paper.\footnote{Note again that an OH-IR object 
does not mean the OH/IR star defined by Habing\,(\cite{hab96}).
The {\it OH-IR objects} considered here include not only evolved stars but also H\,II 
regions, young stellar objects and molecular clouds 
(see column 3 of Table\,1 for details).} 
Among them, 1070 OH-IR objects have IRAS
Low Resolution Spectra (LRS) in the catalog of Kwok et al.\,(\cite{kwo97}).
Chen et al.\,(\cite{che01}) have discussed the statistical properties of a
somewhat smaller sample of 1024 galactic OH-IR objects with LRS
spectra. Note that in that sample there was one object
(\object{IRAS\,16527$-$4001})
counted twice (sequential numbers 222 and 645 in their Table\,1, with entry 645
being the  correct one -- the source belongs to the group A). Also,
\object{IRAS\,16279-4757} with sequential number 778 and
\object{IRAS\,19327+3024} with sequential number 865 in their paper, turned out to be
absorption and non-detection at 1612\,MHz, respectively. The total number of OH-IR 
sources in Chen et al.\,(\cite{che01}) is actually 1021. The present sample of
1070 galactic OH-IR objects that have the IRAS association better than 1\arcmin\ and
have LRS spectra contains 49 new objects. All these 49 new maser sources are listed in
Table\,1a\footnote{Table\,1a is only accessible in electronic form  at
the CDS via anonymous ftp to cdsarc.u-strasbg.fr (130.79.128.5) or via
http://cdsweb.u-strasbg.fr/Abstract.html. The following papers have been cited
in Table\,1a: Chengalur et al.\,(\cite{che93}), Eder et al.\,(\cite{ede88}),
Lewis et al.\,(\cite{lew90}), Lewis\,(\cite{lew92}, \cite{lew94}), Sevenster et
al.\,(\cite{sev97}, \cite{sev01}), Dickinson and
Chaisson\,(\cite{dic73}), Szymczak et al.\,(\cite{szy01a}),
te Lintel Hekkert et al.\,(\cite{tel89})} 
which is an extension of Table\,1 of Chen et
al.\,(\cite{che01}) and has the same structure as their Table\,1. That is to say, 
the sources in Table\,1a are grouped using the University of Calgary LRS classification 
scheme with the letter codes defined by Volk \& Cohen\,(\cite{vol89}) and, within each LRS 
group, ordered by right ascension. The last column of Table\,1a contains information
about references for the OH maser observations. Note that due to better maser
position determination (ATCA3 - Sevenster et al.\,\cite{sev01}), the 4 sources
(\object{IRAS\,18440$-$0020}, \object{IRAS\,18443$-$0147},
\object{IRAS\,18102$-$1828}, and \object{IRAS\,18449$-$0514}), which were denoted by ``1'' 
in the Table\,1 of Chen et al.\,(\cite{che01}) (meaning that the radio and IRAS positions 
differ by more than 1\arcmin and hence these objects were not counted as OH-IR sources), are
now included in Table\,1a as OH-IR objects.

We have searched the ISO Data Archive for SWS observations around 34.6\,$\mu$m taken 
within 1\arcmin\ of the IRAS position for all the 1876 OH-IR objects. 
There are 86 sources with a total of 170 SWS spectra that 
cover the wavelength region of interest. We also added \object{NML\,Cyg} 
to our sample. This object has no IRAS name, but it has 8
associated 34.6\,$\mu$m ISO SWS spectra. Hence the total number of galactic
OH-IR objects considered in this paper is 87 and the number of their ISO SWS
spectra is 178.

The ISO Spectral Analysis Package (ISAP\,2.1)\footnote{The ISO Spectral
Analysis Package (ISAP) is a joint development by the LWS and SWS Instrument
Teams and Data Centers. Contributing institutes are CESR, IAS, IPAC, MPE, RAL
and SRON.} was used to process and analyze the 178 spectra in our sample.
During data reduction, glitches were removed carefully, but small memory effects
were smeared out by direct averaging across the two sub-scans. However, in the case
of large memory effects, the up and down scans were averaged separately to
verify the reality of the 34.6\,$\mu$m absorption feature.

ISO spectra observed in different AOTs or speed modes have different spectral
resolutions. Most of the OH-IR objects in the sample have only ISO SWS\,01
spectra available. The ISO SWS\,01 data were obtained in four speed modes: 1,
2, 3 or 4, with corresponding spectral resolution at 34.6\,$\mu$m of about
400, 500, 800 and 1500, respectively. The remaining spectra in the sample were 
taken with ISO SWS\,02, 06 or 07 modes for which the spectral resolutions 
are about 2250, 1500 and 30000, respectively. Among the 178 analyzed ISO SWS
spectra, 126 are SWS\,01 (with 50 of them obtained in speed 1, 45 in speed
2, 22 in speed 3 and 9 in speed 4 mode), 23 are SWS\,02, 20 are SWS\,06 and 9 
are SWS\,07 spectra.

The details of our search and analysis are given in 
Table\,1\footnote{Table\,1 is only accessible in electronic form  at
the CDS via anonymous ftp to cdsarc.u-strasbg.fr (130.79.128.5) or via
http://cdsweb.u-strasbg.fr/Abstract.html. The following papers have been cited
in Table\,1: Braz et al.~(\cite{bra90}), Chengalur et al.~(\cite{che93}),
David et al.~(\cite{dav93}), Dickinson \& Chaisson~(\cite{dic73}),
Eder et al.~(\cite{ede88}), Engels~(\cite{eng79}), Ivison et al.~(\cite{ivi94}),
Le Squeren et al.~(\cite{les92}), Lewis~(\cite{lew94}), 
Likkel~(\cite{lik89}), Sevenster et al.~(\cite{sev97}, \cite{sev01}),
Silva et al.~(\cite{sil93}),  Szymczak et al.~(\cite{szy01a}), 
te Lintel Hekkert~(\cite{tel90}, \cite{tel91a}),
te Lintel Hekkert \& Chapman~(\cite{tel96}), 
te Lintel Hekkert et al.~(\cite{tel89}, \cite{tel91b}), Wilson \& Barrett~(\cite{wil72}),
Zijlstra et al.~(\cite{zij89}).}
The OH-IR objects in Table\,1 are grouped according to
the detection of 34.6\,$\mu$m absorption line and ordered within each
group by their IRAS name while the spectra belonging to the same OH-IR object 
are ordered  by their TDT number. An object is classified as: group
`A' if it has at least one `a'-type spectrum; group `T' if
it has at least one `t'-type spectrum but no `a'-type spectrum; 
group `U' if it has only `n'-type and at least one 
`u'-type spectra; group `N' if it has only `n'-type spectra.
Our definition of the ISO spectra is given below and described in 
more details in Sect.\,\ref{statisticsp}.

The above classification of the analysed objects is given in column (1) of Table\,1.
The other columns contain the following data: (2) - the IRAS
name; (3) - the class of the object as defined in SIMBAD (http://simbad.u-strasbg.fr/), 
and the abbreviations that are explained in the corresponding ReadMe file; 
(4) - the TDT number (Target Dedicated Time: the number that
identifies an ISO observation); (5) ISO Astronomical Observational Template
(AOT), with number in parentheses showing the scan speed for SWS\,01 spectra;
(6) - a coding (Sp) related to the 34.6\,$\mu$m line in the spectrum (`a':
the absorption at 34.6\,$\mu$m is detected; `t': it is tentatively detected; `u': it is
undetected while we expect that it should be seen; and `n': the spectrum has too
small resolution and/or is too noisy and/or the expected strength of the
34.6\,$\mu$m line is too weak -- the attribution of the code letter for
each spectrum is explained in  Sect.\,\ref{statisticsp});
(7) - the continuum flux around 34.6\,$\mu$m ($F^{\rm c}_{\rm 34.6}$) for each
spectrum; (8) - the 1\,$\sigma$ noise level of the 34.6\,$\mu$m spectrum, which is 
computed as the average of the standard deviations of all the data points in the
range from 34.3 to 34.9\,$\mu$m - the reciprocal of this quantity is
proportional to the signal to noise ratio of the spectrum; (9) - the representative 
34.6\,$\mu$m continuum flux level for each OH-IR object 
($\bar F^{\rm c}_{\rm 34.6}$) - it is derived by averaging mainly among good-quality
SWS\,01 spectra of each object; (10) -
the line center flux density of the 34.6\,$\mu$m feature with the continuum 
base subtracted ($F^{\rm p}_{\rm 34.6}$)
(it is computed only for spectra that show this absorption line and a negative value 
means that the line 
feature is in absorption); (11) - notes related to $F^{\rm p}_{\rm 34.6}$;
(12) - the blue peak flux density of the OH
1612\,MHz maser emission ($F^{\rm p,blue}_{\rm OH}$); (13) and (14) - references
and notes, respectively, for the maser peak fluxes with abbreviations explained 
at the end of the table. More detailed explanations for some of the quantities shown 
in Table\,1 are given below.

\section{The 34.6\,$\mu$m absorption line}
\label{detectionrate}

The detection rate of the 34.6\,$\mu$m line can be affected by at least four
factors: (1) spectrum resolution, (2) noise level, (3) continuum flux level
around 34.6\,$\mu$m and (4) intrinsic line strength. The first two factors
interplay with each other in determining the appearance of any line feature in
the spectrum: for a lower resolution the absorption line is shallower and hence 
a lower noise level is required for the line to be resolved.
The 34.6\,$\mu$m feature is more easily detected when the adjacent
continuum flux  is high, since in this case, for a given observational set-up, the
signal to noise ratio is higher. Finally, the intrinsic 34.6\,$\mu$m
line strength is a critical factor which can override all of the above
factors: for a
weaker line,  a higher resolution and/or higher signal to noise ratio
are required to detect the line. In the next sub-section, we introduce a quantitative
concept of the 34.6\,$\mu$m line detectability that takes into account all
these factors, and we will discuss the effects of each of them on our spectra.

\subsection{The 34.6\,$\mu$m absorption line in the sample spectra}
\label{statisticsp}

\begin{figure}[]
       \centering
\resizebox{\hsize}{!}{\includegraphics[angle=0]{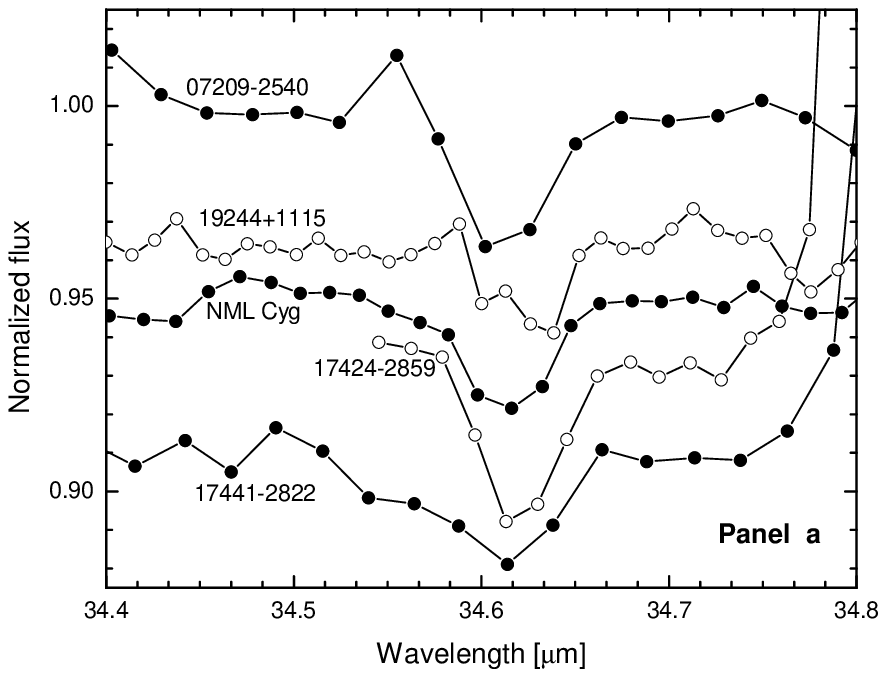}}
\resizebox{\hsize}{!}{\includegraphics[angle=0]{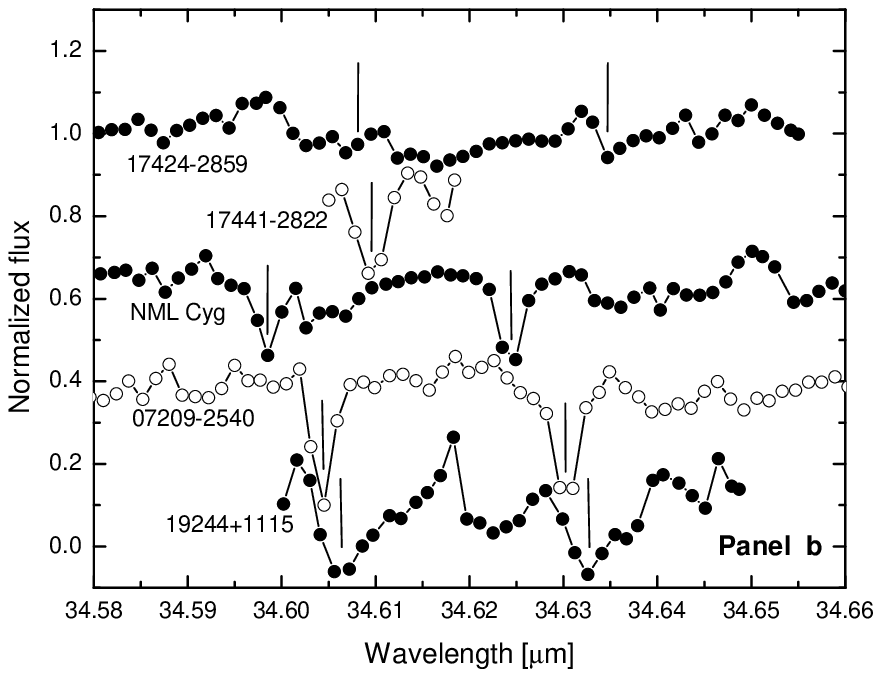}}
       \caption{The representative 34.6\,$\mu$m absorption line profiles for the 5 detection
objects. Panel a shows the best spectra among the SWS\,01, 02 or 06 ones of each object; 
Panel b shows the best SWS\,07 spectrum of each object. The line centers or the expected 
line centers of the 34.6\,$\mu$m doublet in Panel b are marked by short vertical lines. The 
spectra in both panels have been shifted arbitrarily for the sake of convenience. The 
TDT identification of the spectra shown in the figure are as follow: 
\object{IRAS\,07209$-$2540} (73402218, 73601963); \object{IRAS\,17424$-$2859} (09401504, 
69602001); \object{IRAS\,17441$-$2822} (28702002, 46001217); 
\object{IRAS\,19244$+$1115} (36401631, 36401613); \object{NML\,Cyg} (52200719, 52200201).}
       \label{he_fig1}
\end{figure}

There are only 5 objects that show the 34.6\,$\mu$m absorption 
line in the ISO SWS spectra (all of them have been discussed by other authors). 
In Fig.\,\ref{he_fig1}, representative spectra are shown for these 5 objects.
Panel {\it a} of the figure shows the best spectra of SWS\,01, 02 or 06 mode for each 
of them, while panel {\it b} shows their best SWS\,07 spectra (resolution R=30000). 
Note that variations of the wavelength positions of the 34.6\,$\mu$m 
doublet, as clearly seen in panel {\it b} of the figure, are possibly due to the different 
radial velocities of these objects.

The remaining objects, except some tentative detections discussed below, do not show the 
34.6\,$\mu$m absorption line. In order to understand such a high number of 
non-detections, we start with an 
estimation of the expected strength of the 34.6\,$\mu$m line for each spectrum, 
using available observational data. Theory predicts that about four
34.6\,$\mu$m photons are needed to produce one 1612\,MHz maser photon 
 in a radiatively pumped maser (Elitzur\,\cite{eli92}). This can be expressed as
$N^{\rm int}_{34.6}=4\,N^{\rm int}_{\rm OH}$, here the left quantity is the
integrated photon flux of the 34.6\,$\mu$m absorption and the right quantity is the
integrated photon flux of the 1612\,MHz OH maser emission. Although some OH masers in 
our sample may not be radiatively pumped, we still can temporarily assume that they are all radiatively 
pumped so as to define a `pseudo' 34.6\,$\mu$m absorption line strength from the maser strength 
for each of them. These `pseudo' line strengths should be 
quite different from their real 34.6\,$\mu$m line strengths, and hence 
we may be able to use this difference as a diagnostic tool to identify them. 
For a line at a frequency $\nu$, the integrated photon flux $N^{\rm int}$ is related to the integrated
energy flux $F^{\rm int}$ by $N^{\rm int}$ = $F^{\rm int}/h\nu$. The
peak flux density of this line, $ F^{\rm p}$, is related to $F^{\rm
int}$ by
$F^{\rm int}  \approx F^{\rm p}\Delta \nu =
F^{\rm p}\,\Delta V\,\nu /c$
(where $\Delta \nu$ is the line width in Hz, $\Delta V$ is the
line width in velocity range and c is the speed of light). Assuming further 
that the maser emission and IR absorption occur in the same 
volume of gas, the velocity range ($\Delta V$) of the maser line 
and of its pumping lines should be similar. 
Therefore, the OH-maser radiative pumping mechanism expressed by
integrated photon fluxes above can be transformed into a relation
between peak flux
densities: $F^{\rm p}_{34.6}=4\,F^{\rm p}_{\rm OH}$. We
consider only the blue peak fluxes ($F^{\rm p,blue}_{\rm OH}$)
for double peaked OH 1612\,MHz masers because the observed 
absorption at 34.6\,$\mu$m is produced by the near side of
the circumstellar envelope. For sources with a 
single peak maser, we take the only peak as 
the blue one (see the note mark `1p' in column 14 of 
Table\,1. The blue peak OH 1612\,MHz maser emission flux
densities are collected from the literature and presented in column 12 of
Table\,1 for each OH-IR object. 

The measured depth of a line in a real spectrum also depends on the
spectral resolution. It is clear that the more points we have within the line 
profile the better information we have about the line depth. There is 
no general agreement on how many points in a line profile are necessary to 
detect the line clearly. Our further considerations show that the assumption that 
at least 4 sampling points within the line profile are necessary to resolve the line clearly 
is appropriate for our study. In this case, with 
$\Delta V \approx V_{exp} \approx 20$\,km/s for a typical circumstellar 
masing shell,  a resolution  
of $R\approx 45000$ is required at 34.6\,$\mu$m. Therefore, the expected 
peak flux density of the 34.6\,$\mu$m line in a spectrum obtained with 
resolution R can be expressed as:
        $$F^{\rm p}_{\rm 34.6,exp}\,=\,4\,F^{\rm p,blue}_{\rm OH}\,R/45000$$
This formula is valid for almost all of our SWS spectra except the SWS\,07 ones for
which the 34.6\,$\mu$m doublet is clearly resolved into two components. In this
case, the above reasoning is only valid for the estimation of the depth of only
one fine structure component and hence, assuming that the strengths of the two
components are the same, the depth of a single component becomes:
      $$F^{\rm p}_{\rm 34.6,exp}\,=\,2\,F^{\rm p,blue}_{\rm OH}\,R/45000.$$

\begin{figure}[]
       \centering
\resizebox{\hsize}{!}{\includegraphics[angle=0]{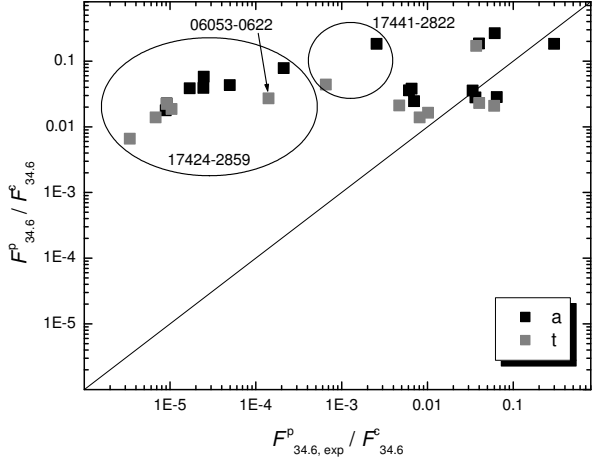}}
       \caption{The 34.6\,$\mu$m absorption line depths measured in the
spectra versus the expected values for these spectra (both divided by
continuum flux $F^{\rm c}_{34.6}$). The location of the spectra of 
\object{IRAS\,17424$-$2859}, \object{IRAS\,17441$-$2822} and 
\object{IRAS\,06053$-$0622} are indicated explicitly. The adopted
classification of the spectra (a,t) is explained in Sect.\,\ref{statisticsp}.}
       \label{he_fig2}
\end{figure}
However, prominent uncertainties may exist in the two formulae. The line widths of 
the 34.6\,$\mu$m absorption and the 1612\,MHz OH maser emission peaks may 
actually be different and both of them may be smaller than the expansion velocity of 
the envelope ($\approx $20\,km/s). These differences may cause changes in 
the coefficients of the two formulae. Thus it is necessary to verify the above 
formulae by comparing (in Fig.\,\ref{he_fig2})  the observed value $F^{\rm
p}_{34.6}$ (given in column 10 of Table\,1) with the relevant 
estimated value $F^{\rm p}_{\rm 34.6,exp}$ for all the spectra in which the 
34.6\,$\mu$m absorption line has been clearly detected. In Fig.\,\ref{he_fig2}, 
both quantities have been divided by the corresponding 34.6\,$\mu$m continuum
flux level ($F^{\rm c}_{34.6}$ given in column 7 of Table\,1) to
remove the effect of different distances and different continuum flux
levels. Excluding the spectra of 
\object{IRAS\,17424$-$2859}, \object{IRAS\,17441$-$2822} and 
\object{IRAS\,06053$-$0622} (see discussion below),
we conclude, by comparison with the one-to-one line, that the {\it expected}
line depths agree in order of magnitude with the measured depths. This
gives us confidence that the above two formulae and their coefficients are good 
enough to be used for the estimation of a reasonable {\it expected} 34.6\,$\mu$m line 
strength. Comparison of the {\it expected} line strength with the noise level 
of the spectrum will allow us to determine whether the absence of the 34.6\,$\mu$m 
absorption line in so many spectra is due to too high spectral noise.

\begin{figure}[]
     \centering
\resizebox{\hsize}{!}{\includegraphics[angle=0]{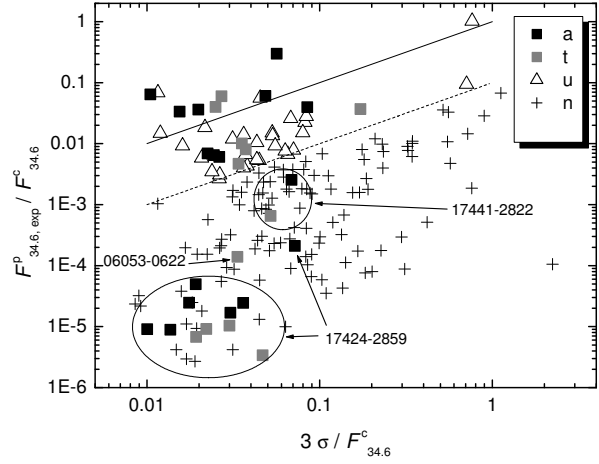}}
       \caption{The expected 34.6\,$\mu$m absorption line depth estimated using
OH 1612\,MHz maser blue peak flux shown against the 3\,$\sigma$ noise level
(both divided by continuum flux $F^{\rm c}_{34.6}$) for all the
sample spectra. The solid and dotted lines mark where the expected 34.6\,$\mu$m 
absorption line depth equals to the 3\,$\sigma$ noise and tenth of the 3\,$\sigma$ 
noise, respectively. The location of the `a' and `t' spectra of 
\object{IRAS\,17424$-$2859},
\object{IRAS\,17441$-$2822} and \object{IRAS\,06053$-$0622} are indicated explicitly.
The adopted classification of the spectra (a, t, u, n) is explained in Sect.\,\ref{statisticsp}.}
       \label{he_fig3}
\end{figure}

Usually a line is considered as detected if its peak intensity is at
least 3 times the root mean square noise level $\sigma$  of the
spectrum. In Fig.\,\ref{he_fig3} we plot $F^{\rm p}_{\rm
34.6,exp}$/$F^{\rm c}_{34.6}$  as a function of 3$\sigma$/$F^{\rm
c}_{34.6}$ for all the spectra of our sample. Squares (black and
grey) correspond to objects for which the 34.6\,$\mu$m absorption
line depth has been derived. The solid line corresponds
to  $F^{\rm p}_{34.6,exp}=3\sigma$. As can be seen from 
the figure, we have been able to detect the 34.6\,$\mu$m absorption line 
in some spectra with $F^{\rm p}_{\rm 34.6,exp}$ significantly
lower than  3$\sigma$ (down to $3\,\sigma$/10, as shown by the filled or 
grey squares above the dotted 
line in the figure). That is to say, in some cases, with an assumed 
maser pumping efficiency of 0.25, the real 34.6\,$\mu$m absorption 
is deeper than the depth required to produce the observed maser flux. 
There are various reasons for this. For example, IR and maser variability 
may cause large uncertainty in the {\it expected} 34.6\,$\mu$m absorption 
depth; the assumption of the same line width $\Delta V$ for both IR and 
maser lines may be inaccurate, although not seriously wrong; 
real masers may transform IR photons into maser photons less effectively 
than assumed in the model of Elitzur\,(\cite{eli92}); 
some OH masers may be hidden behind OH molecular clouds that absorb 34.6\,$\mu$m 
photons but do not mase. The distribution of data points in Fig.\,\ref{he_fig3} 
shows that we can take 
$3\,\sigma$/10 as a conservative limit for the detection of the 
34.6\,$\mu$m line. 

The above considerations allow us to assign a class to each spectrum 
(given in column 6 of Table\,1).
Spectra in which the 34.6\,$\mu$m line is
detected and measured are assigned class
`a' if $F^{\rm p}_{34.6}\,\geq \,3\,\sigma$ (full squares in
Fig.\,\ref{he_fig3}) or  class `t' if
$F^{\rm p}_{34.6}\,<\,3\,\sigma$ ( gray squares in Fig.\,\ref{he_fig3}).
Spectra in which the 34.6\,$\mu$m is not detected are assigned class `u' if
$F^{\rm p}_{\rm 34.6,exp}\,\geq \,3\,\sigma/10$ (open triangles in
Fig.\,\ref{he_fig3}) or class `n' if $F^{\rm p}_{\rm
34.6,exp}\,<\,3\,\sigma/10$ (`+' symbols in Fig.\,\ref{he_fig3}).
In other words, in spectra of class `n', we do not expect 
to detect the 34.6\,$\mu$m absorption because of too low resolution 
or too high noise level or too low continuum flux or too weak 
{\it expected} 34.6\,$\mu$m line strength, even if the relevant  
OH masers are radiatively pumped. In spectra of class `u', one can 
expect to see the absorption on the basis of the radiative maser-pumping model, 
but it is actually not seen (the possible
reasons are discussed in Sect.\,\ref{nondetection}).
The fact that many `n'-type spectra are located far below the dotted
line in Fig.\,\ref{he_fig3} implies that  the 34.6\,$\mu$m feature is not 
detected in these objects most probably due to its intrinsic weakness (their 
{\it expected} line strengths are far smaller than the noise level of the spectra).

The data points corresponding to \object{IRAS\,17424$-$2859} and \object{IRAS\,17441$-$2822} 
and \object{IRAS\,06053$-$0622} are below this limit. In case of the two galactic center (GC)
sources IRAS\,17424$-$2859 and IRAS\,17441$-$2822, it is likely that the observed strong 
features at 34.6\,$\mu$m are the result of absorption by intervening foreground gas in the 
direction of GC (see e.g. Kaifu et al.\,\cite{kai72}, Scoville\,\cite{sco72} or Karlsson et 
al.\,\cite{kar03}). On the other hand, for the SWS observations of IRAS\,17424$-$2859 and 
IRAS\,17441$-$2822 we assigned the OH maser \object{OH\,359.946$-$0.048} and 
\object{OH\,0.667$-$0.035}, respectively. In the so crowded region such association could be 
unphysical and, therefore, we are aware of the fact that the unusual behavior of these sources 
could also be related to the complexity of the galactic center region. In consequences, the
quantities derived for them are probably untrustworthy. The situation of 
IRAS\,06053$-$0622, classified by SIMBAD as an ultra-compact \ion{H}{ii} region, is different.
The 34.6\,$\mu$m absorption seems to be present only in one of its SWS\,06 spectra
(TDT 71101802 - see Table\,1). The feature is clearly present in
{\it all} detectors in the up-scan spectra while the down-scans are relatively noisy
and it is difficult to recognize this weak feature in them. In addition, the
1612\,MHz OH maser peak flux is given in te Lintel Hekkert's PhD
thesis\,(\cite{tel90}) but not in any of the published papers, which may
indicate that the maser flux is not very reliable. Therefore, the large
difference between the estimated and the observed 34.6\,$\mu$m line depth for
this object may be due to the bad quality of the observational data. 

\subsection{Spectra statistics}
\label{dependencesp}

For the convenience of the discussion, the 178 analyzed ISO SWS spectra can be
split into 2 broad families: `at' that merges the class-`a' and class-`t'
spectra, i.e., spectra with the 34.6\,$\mu$m line at least tentatively
detected, and `un'  that merges class-`u' and class-`n' spectra, i.e., 
spectra with the 34.6\,$\mu$m line not detected.

\begin{figure}[]
       \centering
\resizebox{\hsize}{!}{\includegraphics[angle=0]{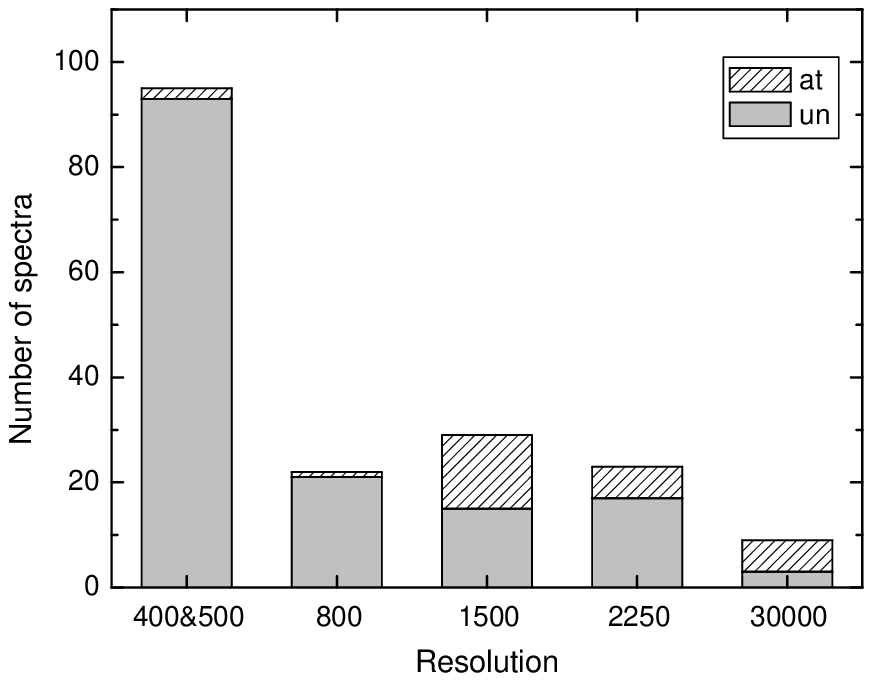}}
       \caption{ Counts of ISO SWS spectra of different resolutions for 
galactic OH-IR objects. The two lowest resolutions are combined
into a single column `400\,\&\,500'. The `at' and `un' type spectra are 
counted separately.}
       \label{he_fig4}
\resizebox{\hsize}{!}{\includegraphics[angle=0]{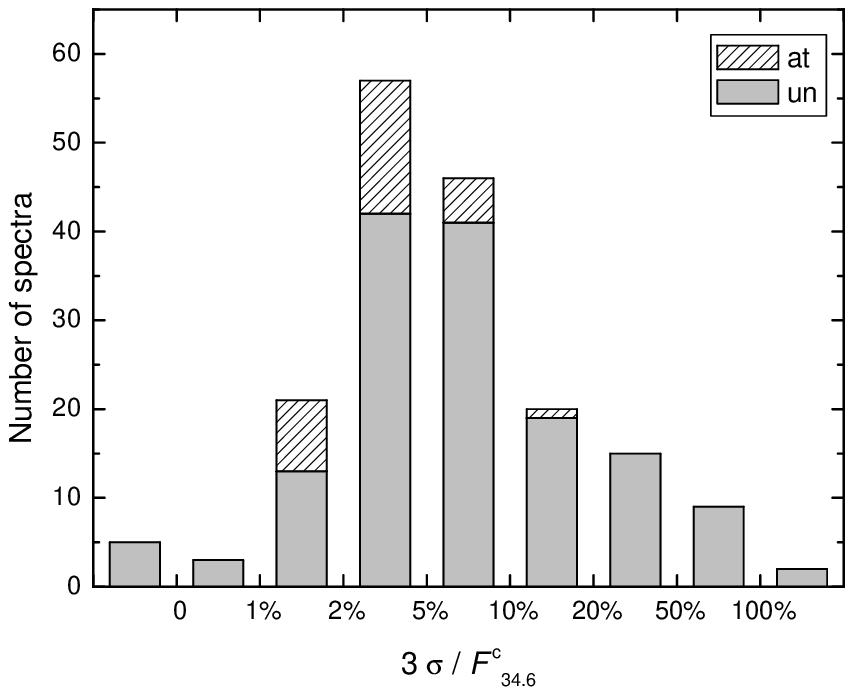}}
       \caption{Counts of ISO SWS spectra of different noise levels 
for galactic OH-IR objects. The negative flux level of the leftmost column 
is due to wrong dark subtraction from the spectra, the effect not corrected 
during our analysis. The `at'  and `un' type spectra are counted separately.}
       \label{he_fig5}
\end{figure}

Fig.\,\ref{he_fig4} illustrates how the detection rate of the  34.6\,$\mu$m line
depends on the resolution. The hatched and grey areas in the stack
columns represent the number of `at' and `un' spectra, respectively, 
and each column corresponds to a single resolution, except resolution  400 and
500 which are combined into a single column `400\,\&\,500'. As  seen,
most of the spectra are taken with low resolution and, in consequence, the
detection of a relatively narrow line feature, such as the 34.6\,$\mu$m one, is
difficult. The detection rate of the 34.6\,$\mu$m feature is very
small (about 2\,\%) for the low resolution bins 400\,\&\,500 and 800,
and much larger for higher resolution bins (30 -- 67\,\%). We call this
dependence the {\em resolution selection effect} on the detection rate of the
34.6\,$\mu$m absorption line. Note that, even in the highest
resolution bin, the detection rate is far from 100\%. This strongly
suggests that  factors other than just the lack of sufficient resolution 
play a role in the fact that the line is not detected in many objects.

Fig.\,\ref{he_fig5} illustrates how the detection rate of the  34.6\,$\mu$m line
depends on the noise level. The hatched and grey areas in the stack
columns now represent the number of `at' and `un' spectra, respectively,
obtained in different intervals of 3\,$\sigma$/$F^{\rm c}_{34.6}$. It 
is seen that `at'  spectra are found only for $3\,\sigma/F^{\rm 
c}_{34.6}\,<\,20\%$. This is what we call the {\em sensitivity selection effect}.
However, a significant fraction of non-detections occurs among 
spectra of intermediate signal-to-noise ratio and, most interestingly, the 
34.6\,$\mu$m line is not detected in any of the spectra with the highest 
signal-to-noise ratio (3$\sigma /F^{\rm c}_{34.6}<1\,\%$). This is due to
the genuine weakness of the 34.6\,$\mu$m line itself as can be
inferred from Fig.\,\ref{he_fig3}.

Globally, the statistics of the detection of the 34.6\,$\mu$m line in 
the 178 analyzed spectra is the following: there are
17 `a' spectra, 12  `t' spectra and 29  `u' spectra. Note that the 
majority of spectra (120) are  of `n' type. Thus,
 excluding the latter, we derive a meaningful detection-rate of 
(17+12)/(17+12+29)=$50\%$.

\subsection{Object statistics}
\label{statisticobj}

We now turn to the discussion of objects statistics, grouping the 
sources as explained in Sect. 2.
    From Table\,1 we see that only 5 galactic OH-IR objects definitely 
show the 34.6\,$\mu$m absorption line (`A' sources).  These  are 
three red supergiants:
\object{NML\,Cyg}, \object{IRAS\,19244$+$1115} $=$ \object{IRC$+$10420} and
\object{IRAS\,07209$-$2540} $=$ \object{VY\,CMa} and two sources related to the
galactic center: \object{IRAS\,17424$-$2859} and
\object{IRAS\,17441$-$28225}. All of them were 
already known to have the 34.6\,$\mu$m absorption feature.
There are two more sources
(\object{IRAS\,06053$-$0622} -- an ultra-compact \ion{H}{ii} region and
\object{IRAS\,17004$-$4119} -- a Mira type) for which the 34.6\,$\mu$m 
feature is tentatively detected (`T' sources).

In summary, out of 87 analyzed OH-IR objects, there are 5 with 
definite 34.6\,$\mu$m detection, 2 with tentative detection, and 16 
in which the detection of the 34.6\,$\mu$m absorption should be possible
according to our predictions, but actually was not. In all the remaining objects, 
the strength of the 
line estimated as explained in Sect. 3.1 is too low to allow detection 
in the analyzed spectra.
  Thus, the detection rate among objects with good enough spectra is 
(5+2)/(5+2+16)=30\%.

\begin{figure}[]
     \centering
\resizebox{\hsize}{!}{\includegraphics[angle=0]{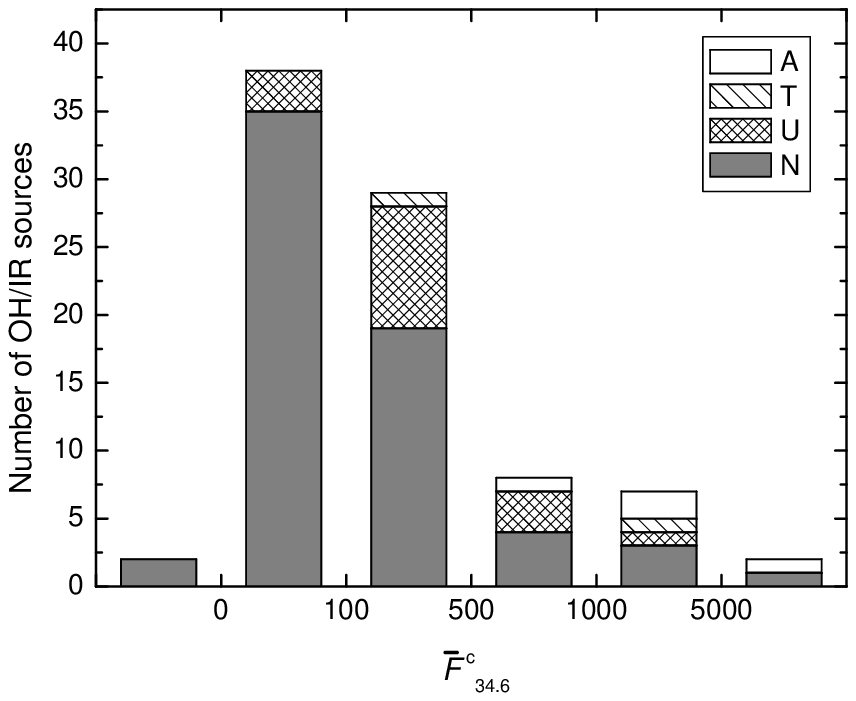}}
       \caption{Counts of sources in 5 intervals of the mean continuum
flux around 34.6\,$\mu$m. 
The leftmost bin corresponds to the sources which have only spectra
with wrong dark current subtraction.}
       \label{he_fig6}
\resizebox{\hsize}{!}{\includegraphics[angle=0]{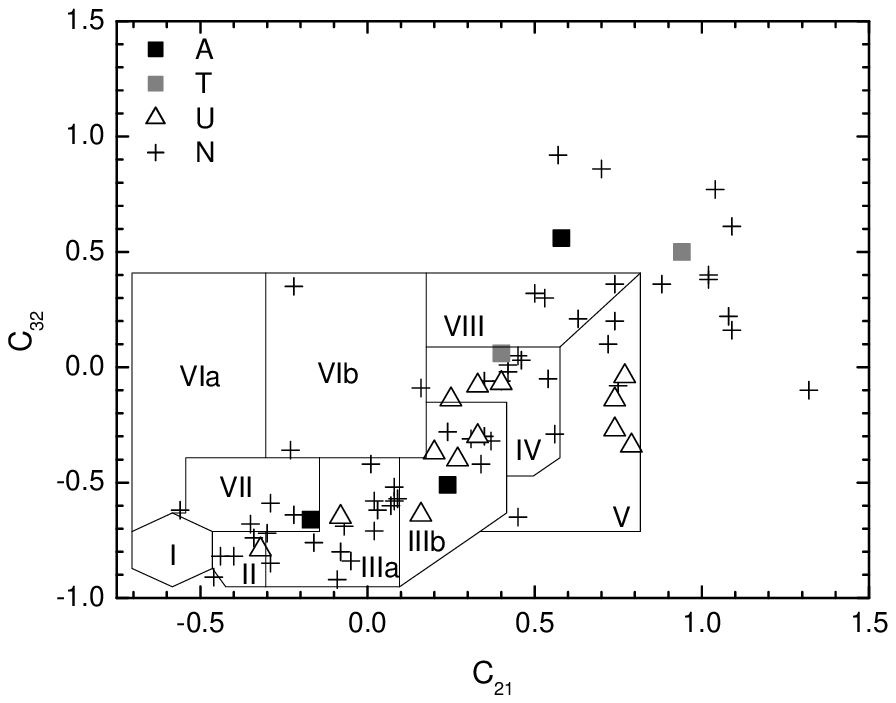}}
       \caption{The IRAS two color diagram for all our sample OH-IR objects with
good quality fluxes at 12, 25 and 60\,$\mu$m. The boxes marked on the plot are
after van der Veen \& Habing\,(\cite{van88}).}
       \label{he_fig7}
\end{figure}

  Of course, the detection rate might be biased by some observational 
effects linked to the properties of the source itself. The sample of sources we 
considered were chosen by different ISO observers for quite different 
observation purposes and is certainly not complete 
or uniform in any sense. For example, the incompleteness in the
brightness distribution of the OH-IR objects can be a potential selection
effect ({\em brightness selection effect}). And the infrared color distribution
of the sample sources may be also incomplete ({\em color selection effect}).

Fig.\,\ref{he_fig6} illustrates how the detection rate of the  34.6\,$\mu$m line
depends on the continuum level. The white, hatched, crossed and grey 
surfaces  in the stack
columns represent the number of `A', `T',`U' and `N' sources
  in different intervals of ${\bar F^{\rm c}_{34.6}}$  
(from column (9) of Table\,1).
  Sources with wrong dark subtraction in their spectra are placed in the first bin 
($\bar F^{\rm c}_{34.6}< 0$). As already mentioned, $\bar F^{\rm c}_{34.6}$ was 
obtained by averaging the 34.6\,$\mu$m continuum fluxes of mainly the
SWS\,01 spectra. However, for some sources, SWS\,01 data are not available and
spectra obtained in other modes were used to estimate the continuum flux.
They are: \object{IRAS\,03507$+$1115} (SWS\,06 -- TDT 80900805), \object{IRAS\,17441$-$2822}
(SWS\,07 -- TDT 46001217), \object{IARS\,17462$-$2845} (SWS\,02 -- TDT 50601406) 
and \object{IRAS\,18349$+$1023} (SWS\,02 -- TDT 52302238).
As Fig.\,\ref{he_fig6} shows, the number
of `A' or `T' sources increases from
0\,\% for the first bin up to 50\,\% for the bin
representing the brightest sources, indicating the presence of a 
{\em brightness selection effect}. The fact that even among the brightest 
sources the detection rate is not 100\% implies that this brightness 
selection effect  cannot account for all cases in which the absorption at
34.6\,$\mu$m is missing.

Now, let us investigate the position of our sources in the 
IRAS color-color diagram to check for a possible {\em color selection effect}. 
The $C_{32}$\,=\,log($F_{60}/F_{25}$) versus 
$C_{21}$\,=\,log($F_{25}/F_{12}$) color-color diagram for
all OH-IR objects from our sample with good quality IRAS fluxes is shown in
Fig.\,\ref{he_fig7}. For reference the regions defined by van der Veen \&
Habing\,(\cite{van88}) are shown. As we can see, the sample sources are 
distributed in all areas where OH-IR objects may appear and there is no clear
tendency of gathering for any type of sources. Therefore, we conclude that 
a {\em color selection effect} is not a severe problem in our sample.

\section{The 1612\,MHz OH maser pumping}
\label{OHpump}

\subsection{The pump rate for type-A sources}
\label{pumprate}

For the five `A' type OH-IR objects, we can estimate a pump rate assuming that
they are radiatively pumped. However, for the two GC sources, 
the physical association between IR sources 
and OH masers in the so crowded region as galactic center may be unphysical and,
in addition, the observed 34.6\,$\mu$m absorption may be due to absorption by 
intervening foreground gas (see discussion in Sect.\,\ref{statisticsp}). Therefore,
we confine the discussion to the three supergiants. 

In Table\,\ref{LineOHpump} we list all the spectra
(column 2) with detected 34.6\,$\mu$m line in the same order as in 
Table\,1. In column (3) we present the measured integrated flux of the 
34.6\,$\mu$m feature ($F^{\rm int}_{34.6}$) in $10^{-19}\,$W/cm$^2$ (`-' sign means 
that the feature is in absorption), while in column (4) we give 
the equivalent widths ($EW_{34.6}$) in $10^{-4}\,\mu$m. These data are used in the 
comparison with the 
EWs of the 53.3\,$\mu$m pumping line given in  He \& Chen\,(\cite{HeCh04}).
For each source with multiple spectra, we choose the one with the highest spectral resolution 
(SWS\,07) among the spectra of good quality. 
The best spectrum is underlined 
in Table\,\ref{LineOHpump} for each source and is used for the pump rate calculation below. 
Also included in Table\,\ref{LineOHpump} is the integrated blue 
peak flux of the OH 1612\,MHz masers ($F^{\rm int,blue}_{\rm OH}$) in 
$10^{-26}$\,W/cm$^2$ together with the corresponding references, in columns (5) and (7), 
respectively. The radiative pump rate defined 
as the ratio of integrated maser emission to integrated IR absorption  
($F^{\rm int,blue}_{\rm OH}\nu_{34.6}/F^{\rm int,best}_{34.6}/\nu_{\rm OH}$ ) 
is presented in column (6).
Column (8) gives the integrated line flux at 34.6\,$\mu$m from the literature 
and column (9) shows the corresponding reference. Our line parameters are found to generally 
agree with those from the literature.

%**************************** Table 2 ****************************
\begin{table*}[]{}
\setcounter{table}{1}
\centering
\caption[]{The 34.6\,$\mu$m and 1612\,MHz line parameters for the three supergiants
which are used to estimate the OH maser radiative pump rate. 
The integrated fluxes $F^{\rm int}_{34.6}$ is in [$10^{-19}$\,W/cm$^2$], 
$F^{\rm int,blue}_{\rm OH}$ is in [$10^{-26}$\,W/cm$^2$] while the equivalent
widths $EW_{34.6}$ is in [$10^{-4}\,\mu$m]. Negative
values mean that the line is in absorption.}
\label{LineOHpump}
\begin{tabular}{l@{ }l@{  }l@{  }l@{ }l@{  }l@{  }l@{  }l@{  }l@{}}
\hline
\noalign{\smallskip}
name &TDT$^{*}$  &$F^{\rm int}_{34.6}$ &$EW_{34.6}$ &$F^{\rm int,blue}_{\rm OH}$ &pump rate &references1 &$F^{\rm int,lit}_{34.6}$ & references2$^{**}$ \\
(1)       &(2)        &(3)                  &(4)         &(5)                         &(6)       &(7)         &(8)                      &(9)            \\
\noalign{\smallskip}
\hline
\noalign{\smallskip}
IRAS\,07209$-$2540~~~    &73103039~~~               &$-18.7~~~      $       &$-18.0~~~      $        &$1751~~~        $ &$0.052 ~~~ $ &tL-H(fig)$^{\#\#}$~~~     &-21.0~~~               &NEU99~~~   \\
                         &73402218               &$-21.4      $       &$-20.9      $        &               &          &              &                    &        \\
                         &$\underline{73601963}$ &$\underline{-17.6}$ &$\underline{-16.2}$  &               &          &              &                    &        \\
IRAS\,19244$+$1115    &31600936               &$-4.12      $       &$-12.6      $        &$662         $ &$0.042$   &SYL97         &-4.8                & SYL97  \\
                         &$\underline{36401613}$ &$\underline{-8.10}$ &$\underline{-21.0}$  &               &          &              &                    &        \\
                         &36401631               &$-3.31      $       &$-10.3      $        &               &          &              &                    &        \\
                         &72400804               &$-6.49      $       &$-15.9      $        &               &          &              &                    &        \\
NML\,Cyg         &05200726               &$-34.7      $       &$-54.4      $        &$1154        $ &$0.052  $ &priv(fig)$^{\#\#}$      &                    &        \\
                         &34201475               &$-8.74^{\#} $       &$-14.7^{\#} $        &               &          &              &                    &        \\
                         &$\underline{52200201}$ &$\underline{-11.5}$ &$\underline{-16.0}$  &               &          &              &                    &        \\
                         &52200719               &$-8.34      $       &$-16.8      $        &               &          &              &                    &        \\
                         &52200720               &$-5.01      $       &$-9.34      $        &               &          &              &                    &        \\
                         &53001311               &$-5.86      $       &$-11.0      $        &               &          &              &                    &        \\
                         &74103105               &$-34.8      $       &$-49.3      $        &               &          &              &                    &        \\
              \noalign{\smallskip}               
\hline 
\end{tabular}
\begin{list}{}{}
\scriptsize
\item[*:] The best spectrum together with its parameters are underlined for each source. 
  They are used to estimate the pump rate.
\item[**:] The reference codes in column (9):
NEU99: Neufeld et al.\,(\cite{neu99});
SYL97: Sylvester et al.\,(\cite{syl97});
\item[\#] Only the sub-line no.10 of 34201475 
spectrum are used.
\item[\#\#] 
            (fig) means the integrated 1612\,MHz OH maser fluxes are measured by us 
            from the figure in the literature.\\
\end{list}
\end{table*}

From Table~\ref{LineOHpump}, one can see that the pump rate is similar
for all the three RSGs and is close to 0.05.\footnote{A formal estimation of the pump rate for
the two GC sources (IRAS\,17424$-$2859 and IRAS\,17441$-$2822)
 gives very small values of $\approx$0.00007 
(estimated from spectrum TDT\,09500706: $F^{\rm int}_{34.6}=$ $-27.7\,10^{-19}$ [W\,cm$^{-2}$]
and data for OH\,359.946$-$0.048: $F^{\rm int,blue}_{\rm OH}=3.51\,10^{-26}$\,[W\,cm$^{-2}$]), 
and $\approx$0.003 
(estimated from spectrum TDT\,46001217: $F^{\rm int}_{34.6}=$ $-1.1\,10^{-19}$ [W\,cm$^{-2}$] 
and data for OH\,0.667$-$0.035: $F^{\rm int,blue}_{\rm OH}=$ $6.0\,10^{-26}$ [W\,cm$^{-2}$]), 
respectively. On the other hand, it is worth to notice 
that the 'pseudo' pump rate for megamaser Arp\,220 is close
to 5! (estimated from spectrum TDT\,11000803: $F^{\rm int}_{34.6}=$ $-1.6\,10^{-19}$ 
[W\,cm$^{-2}$] and its single peak OH 1612 MHz data from Baan \& Haschick\,(\cite{baa87}): 
$F^{\rm int}_{\rm OH}=5700\,10^{-26}$\,[W\,cm$^{-2}$]).}
The similar pump rates for the three RSGs may indicate that their OH 1612\,MHz masers are 
radiatively pumped by the 34.6\,$\mu$m line. A more complete discussion of the pumping 
mechanism in the case of RSGs is presented in He \& Chen\,(\cite{HeCh04}), where the 
contribution of the 53.3\,$\mu$m line is analysed. 

If we adopt 0.05 as the pump rate of a typical radiatively pumped 1612MHz OH maser, we will 
be in contradiction to the theoretic estimation that four 34.6\,$\mu$m photons are needed 
to produce one 1612\,MHz maser photon (a pump rate of 0.25). On the one hand, the too small 
pump rates (1/5 of the theoretic value 0.25) caution that Elitzur's model may be 
too simple to give a proper pump rate because many important processes (e.g. collisional 
effects, line overlap) are still absent in his model. On the other hand, the IR absorption 
line may be wider than the maser emission line and both lines may be narrower than the 
expansion velocity of the circumstellar envelope, and our assumptions (discussed in 
Sect.\,\ref{statisticsp})
can be responsible for a large difference between theoretical and the estimated pump rate.
Note, however, that the two formulae used in this paper were more firmly justified by the 
one-to-one relation check shown in Fig.\,\ref{he_fig2}.

\subsection{Why do we not see the 34.6\,$\mu$m absorption in `U' type sources?}
\label{nondetection}

As we have discussed in Section\,\ref{statisticsp}, there are 16 OH-IR objects 
of which the {\it expected} 34.6\,$\mu$m absorption line depth is large enough to be 
detected in at least some of their ISO SWS spectra but was not detected. In this section 
we discuss possible reasons for these non-detections.

The 34.6\,$\mu$m photons are considered to be the
main pumping mechanism for the OH 1612\,MHz maser {\it only} in late
type stars.
The situation in other galactic sources (e.g. molecular clouds or \ion{H}{ii}
regions) where the maximum of IR flux is shifted to much longer wavelengths is
still far from being completely understood and it is quite possible that the OH
maser pumping by longer wavelength IR photons (e.g. the 53.3\,$\mu$m photons)
could be dominant. The possibility of 1612\,MHz OH maser pumping in
these sources by the 53.3\,$\mu$m photons is discussed in more detail in
another paper (He \& Chen\,\cite{HeCh04}) devoted to the OH 
lines in the ISO Long Wavelength Spectrometer wavelength range. 

Secondly, the non-detections could be related to the non-uniformity of the OH shell.
Assuming that the 34.6\,$\mu$m continuum
emitting dusty medium is located mostly inside  the OH maser shell, the absence
of absorption at 34.6\,$\mu$m may be explained 
by the clumpy nature of the OH shell. Dust grains effectively emitting
34.6\,$\mu$m photons have a temperature above 100\,K. For a typical 
circumstellar envelope of an AGB star, the dust temperature usually decreases 
from about 1000 K from the inner edge to about 100 K in the 
OH molecular shell. Hence the major parts of the 34.6\,$\mu$m emitting 
region are located inside the OH shell and most of the 34.6\,$\mu$m photons 
to pump the OH maser come from the hottest central region close to the central star.
Therefore, it is possible that a sparsely distributed clumpy 1612\,MHz OH maser
shell obscures only a small part of the dust emitting region or even fails to
cover the brightest central area of the IR emitting region. In such a
situation, even if the number of 34.6\,$\mu$m photons absorbed in the individual 
OH clumps is sufficient to pump the observed OH maser, the OH absorption 
efficiency integrated over the whole observable IR emitting area towards the observed
source may be too small to produce an absorption line strong enough to be
detected in the ISO SWS spectra. 
High spatial resolution interferometric observations (e.g. VLBI, MERLIN) of
OH 1612\,MHz masers show that the distribution of maser spots is not circular
and sometimes delineates strip--like regions originating from the star (e.g.
Richards et al.\,\cite{ric99}, Zijlstra et al.\,\cite{zij01}, Szymczak \&
Richards\,\cite{szy01}, Gledhill et al.\,\cite{gle01}, Bains et
al.\,\cite{bai03}). We are aware that due to strong beaming effects in maser
amplification the apparent distribution of maser spots does not reflect the
distribution of masing material adequately (see, e.g. Sobolev et al.\,\cite{sob98}).
However, the spread of maser spots delineates the region where significant absorption by 
the masing gas appears (Sobolev et al.\,\cite{sob04}). Hence, interferometry
of masers shows that the region masing at 1612\,MHz does not encircle the
whole IR emitting region of the star envelope. 
The spatial-velocity structure of OH maser clusters in
OH/IR stars also shows that the distribution of the masing gas is likely to be
non-spherically symmetric and traces the interface regions of the outflows that
are known to be inhomogeneous (e.g., Zijlstra et al.\,\cite{zij01}). Furthermore, 
high-resolution infrared imaging observations of dust emission 
from AGB or post-AGB stars also support the hypothesis of a clumpy nature 
of the circumstellar envelope
(e.g., Goto et al.\,\cite{got02}, Monnier et al.\,\cite{mon04}). Therefore, 
there are numerous pieces of evidence that the gas that mases at 1612\,MHz does not form a
complete IR screen surrounding the star. This makes the inhomogeneity of the OH
"shell" a possible explanation for the large fraction of non-detections of the 34.6\,$\mu$m 
OH absorption line.

Thirdly, it is possible that for relatively extended OH maser shells the limb-filling 
emission effect will play a role. A good example of this effect was
demonstrated by modeling of the galactic center source \object{Sgr\,B2} by
Goicoechea \& Cernicharo\,(\cite{goi02}). They set up a model by placing an
infrared emitting sphere inside a surrounding OH shell. They found
from their modeling that if the OH shell is too far away from the inner sphere that emits 
most of the infrared radiation or if OH shell itself is too geometrically thick, the observed
OH infrared absorption lines will disappear due to the limb-filling emission effect. However, 
it is not obvious whether this model can be adopted for stellar OH-IR objects.
On the other hand, if the OH shell is detached from the main body of the
central infrared emitting region, the above scenario should be possible.

It is still tempting to consider collisional excitation of OH molecules as the main
reason responsible for frequent non-detections of the 34.6\,$\mu$m absorption
line. However, collisional OH maser pumping has been considered as rather inefficient
since the microwave OH maser and the infrared emission are observed to co-vary
in most stellar OH-IR objects (Harvey et al.\,\cite{har74}). An additional argument
against this mechanism (at least in the case of \object{Sgr\,B2}) comes from the
ISO observations of  OH $\Lambda$-doublets. These observations do not show any
asymmetries in the line intensities of these infrared doublets while they are expected in
the collisional pumping process (see Goicoechea \& Cernicharo\,\cite{goi02} and
references therein). One may argue that OH masers other than late type stars 
may exist in our sample of OH-IR objects and hence collisional pumping can 
be responsible for the missing of the 34.6\,$\mu$m line in some sources. However,
according to the SIMBAD classification, among the 16 type-U 
objects {\it only one} is classified as a molecular cloud while the other 15 are {\it
stellar objects}. Therefore, it seems that collisional pumping cannot explain the 
non-detections discussed here.

The light variation in the IR and/or microwave ranges cannot be
a major factor for the explanation of the non-detections. If sources are in
minimum light during the ISO observation, their IR emission and hence
radiative maser pumping process may be weaker, while on the other hand, 
if the relevant OH maser emission is in maximum light during the maser 
spectroscopic measurement (which is not simultaneous with the ISO measurement), 
the measured maximum maser flux will result in the estimation of a strong enough 
{\em expected} 34.6\,$\mu$m absorption strength. In such a case, we are  
comparing a maximum {\em expected} 34.6\,$\mu$m line strength with a minimum 
observed strength and non-detection (as for our type `U' sample sources) may occur.
 However this effect probably cannot explain all of our 16
type `U' sources because the IR light variations are usually small and it is unreasonable to
assume that all of them were in minimum light at the time of the ISO observations and 
were in maximum light at the time of maser observations. 

\section{Summary}
\label{summary}

To search for the pumping line of the OH 1612\,MHz masers we have
analyzed 178 ISO SWS spectra taken around 34.6\,$\mu$m for 87 OH-IR objects. 
The performed analysis shows that the 34.6\,$\mu$m line is usually weak in 
our sample of OH-IR objects and the signal-to-noise ratio and spectral resolution of 
120 spectra (among 178) are not high enough to resolve this feature. 
Among the 23 OH-IR objects associated with the remaining 58 spectra, 7 objects
are of type `A' or `T' (the 34.6\,$\mu$m line is detected or tentatively detected) and 
16 objects are of type `U' (the 34.6\,$\mu$m line is expected to be seen but
it is not seen). Therefore, we obtain about $30\%$ (7 out of 23) detections and 
about $70\%$ (16 out of 23) non-detections. We have discussed some possible explanations 
for the 16 non-detections among the 23 OH-IR objects with reliable spectra, including FIR 
photon pumping, sparse distribution of the 1612\,MHz maser spots and the limb-filling emission
effect that could occur in a large OH shell.

The 34.6\,$\mu$m absorption line is found only in three RSGs and two galactic 
center sources.   
The pump rates of red supergiants,
calculated assuming that their 1612 MHz OH masers are radiatively pumped by this 
IR line, turned out to be similar and close to 0.05, a value which is only 1/5 of the 
theoretical value of 0.25. The similar pump rates of the 3 RSGs indicate that they might all 
be radiatively pumped. 

\begin{acknowledgements}
We thank the ISO helpdesk for their earnest help in solving our problems in
installing and using ISO data processing software. We are also indebted to Dr.
Albert Zijlstra for his comments, to Dr. Gra\.zyna Stasi\,nska for inspiring discussions and 
help during preparation of the manuscript, to the referee Dr. A. Winnberg and to the A\&A Editor 
M. Walmsley for the valuable comments improving the original manuscript. Our work also benefited 
from the government scientific cooperation joint project between China and Poland for the years
2001-2003 and is supported by the Chinese National Science Foundation under
Grant No. 10073018, the Chinese Academy of Sciences Foundation under
Grant KJCX2-SW-T06 and the Yunnan Natural Science Fund (2002A0021Q). AMS
acknowledges financial support by RFBR (grant 03-02-16433) and the Ministry of
Industry, Science and Technology of the Russian Federation (Contract
No.\,40.022.1.1.1102). This work has also been partly supported by grant
2.P03D.017.25 of the Polish State Committee for Scientific Research and 
by the European Associated Laboratory "Astrophysics Poland-France".
\end{acknowledgements}

\end{document}